\documentclass[10pt,fleqn]{article}
\usepackage{amsmath}
\usepackage{amssymb}
\usepackage{graphicx}
\usepackage{epsfig}
\usepackage{subfigure}
\usepackage[margin=1.40cm,nohead]{geometry}
\usepackage{multicol}

\makeatletter

\newenvironment{figurehere}
  {\def\@captype{figure}}
  {}
\makeatother

\begin{document}
\title{Fourier expansions for the potentials of lattices of charge}
\author{Jonathan Landy\\
\small Physics and Astronomy Building, \\ \small University of
California
Los Angeles, Los Angeles, CA 90095-1547\\
\small \texttt{landy@physics.ucla.edu}}

\date{\today }

\maketitle
\begin{multicols}{2}

In this note we apply the Poisson sum rule \cite{CRC-96} to obtain formal expressions for the Fourier coefficients of the potential of a lattice of generalized charge.  Each generalized charge is assumed to contribute to the potential a term which depends only on the vector displacement from the charge's position.  The coefficients are explicitly calculated for two types of individual charge potentials: Coulomb and Yukawa.  For one and two-dimensional Coulomb lattices, we find that the non-zero frequency components decrease exponentially both with the distance to the distribution and with frequency.  The exponential convergence of these expansions indicates that a truncated Fourier series will often provide both a simple and accurate analytic approximation for these potentials.  In particular, this result may often be applied to justify use of the continuous charge distribution approximation for observation points further from the distribution than the spacing between the charges.  This result also serves to explain numerical observations which have been presented previously \cite{Par-75,Vau-92,Vau-00}.  The resulting expansions for lattices of Yukawa type charges are seen to be closely related to those of their Coulomb analogs.  Yukawa potentials are physically realized in linearly screened Coulomb systems \cite{Han-00}, for example, and so are of direct physical significance.   We end the note with a brief study of finite and disordered lattices.

The general form of the Poisson sum rule allows for a sum over a periodic lattice to be replaced by an integral over the volume of the lattice.  For an arbitrary individual particle potential function $f$, the rule states
\begin{eqnarray}
\label{Poisson}
\sum_{\textbf{R}} f(\textbf{r}-\textbf{R}) = \frac{1}{V} \int \sum_{\textbf{G}}f(\textbf{r}-\textbf{R}) \exp(i \textbf{G}\cdot \textbf{R}) d^n R.
\end{eqnarray}
Here, $n$ is the dimension of the lattice, $V$ is the $n$-dimensional volume per unit cell of the direct lattice, the vectors $\textbf{R}$ are the direct lattice vectors of the distribution, and the vectors $\textbf{G}$ are the reciprocal lattice vectors of the distribution \cite{Ash-76}.  Shifting the origin of integration in Eq.(\ref{Poisson}) to the projected position of the observation point $\textbf{r}$ immediately gives formal integral representations for the Fourier coefficients of the potential.  

For a lattice of Coulomb point charges of charge $q$, the individual potential function is given by 
\begin{eqnarray}
f(\textbf{r}) = \frac{q}{|\textbf{r}|}.
\end{eqnarray}
The evaluation of the right side of Eq.(\ref{Poisson}) for the general two-dimensional Coulomb lattice is carried out in the appendix.   In each dimension the integrals may be carried out readily and we find the following expansions for the potentials of one, two, and three-dimensional Coulomb lattices, respectively.
\begin{eqnarray}
\label{oned}
\Phi_{C,1} &=& \bar{\Phi}_{C,1} +  2 \lambda \sum_{\textbf{G}\not = 0}  K_0(|\textbf{G}|d)\exp[i \textbf{G}\cdot \textbf{r}], \\ 
\label{twod}
\Phi_{C,2} &=& \bar{\Phi}_{C,2} + 2 \pi \sigma \sum_{\textbf{G} \not = \textbf{0}} \frac{1}{|\textbf{G}|} \exp[i \textbf{G}\cdot \textbf{r}-|\textbf{G}|d], \\
\label{threed}
\Phi_{C,3} &=& \bar{\Phi}_{C,3} + 4 \pi \rho \sum_{\textbf{G} \not = \textbf{0}} \frac{1}{|\textbf{G}|^2} \exp[i \textbf{G}\cdot \textbf{r}].
\end{eqnarray}
Here, $\bar{\Phi}_{C,n}$ is the potential one would obtain through the appropriate continuous charge distribution approximation, $d$ is the perpendicular distance to the distribution, and the function $K_0$ is a modified Bessel function of the second kind.  For large arguments, the function $K_0$ may be expanded as \cite{Abr-72}
\begin{eqnarray}
K_0(z) \sim \sqrt{\frac{\pi}{2 z}}e^{-z} \sum_{j=0}^{\infty}\frac{(-1)^j \prod_{k=0}^{j}(2k+1)^2}{j! (8z)^j}.
\end{eqnarray}
It follows that the non-zero frequency components will indeed be exponentially damped with both the frequency and the distance from the distribution for one and two-dimensional Coulomb lattices.  The three-dimensional expansion Eq.(\ref{threed}), while not exponentially convergent, also converges at a faster rate than the original Coulomb sum.  In all three cases we observe that potential oscillations perpendicular to high density lattice planes dominate those oscillations perpendicular to low density lattice planes.   We note that the one and three-dimensional Coulomb results Eq.(\ref{oned}) and Eq.(\ref{threed}) have been discussed previously and have been applied to obtain expressions for a lattice's Madelung constant  \cite{Hau-75, Cra-87}.  The general two-dimensional result Eq.(\ref{twod}) appears to have escaped prior notice, however.

The integrations required to obtain the potentials of lattices of Yukawa type charges, with individual particle potentials of the form $q \exp[-\gamma r]/r$, may also be carried out, though the evaluations are slightly more difficult.  In the context of linearly screened Coulomb systems, the parameter $\gamma$ is proportional to the square root of the concentration of screening counterions.  The Fourier expansions obtained through application of Eq.(\ref{Poisson}) for the potentials of one, two, three-dimensional Yukawa lattices are
\begin{eqnarray}
\label{oneds}
\Phi_{1,Y} &=& 2 \lambda \sum_{\textbf{G}}  K_0((|\textbf{G}|^2 + \gamma^2)^{1/2}d)\exp[i \textbf{G}\cdot \textbf{r}], \\ 
\label{twods}
\Phi_{2,Y} &=& 2 \pi \sigma \sum_{\textbf{G}} \frac{ \exp[i \textbf{G}\cdot \textbf{r}-(|\textbf{G}|^2 + \gamma^2)^{1/2}d]}{(|\textbf{G}|^2 + \gamma^2)^{1/2}}, \\
\label{threeds}
\Phi_{3,Y} &=&  4 \pi \rho \sum_{\textbf{G}} \frac{1}{|\textbf{G}|^2+\gamma^2} \exp[i \textbf{G}\cdot \textbf{r}].
\end{eqnarray}
It is interesting to observe that the non-zero frequency coefficients above may always be obtained from the corresponding unscreened coefficients by replacing  $|\textbf{G}|$ by $(|\textbf{G}|^2 + \gamma^2)^{1/2}$.  It is also interesting that the screened systems' zero-frequency coefficients always have a mathematical form similar to their non-zero frequency coefficients.  The Coulomb expansions may be obtained from these results by taking the limit $\gamma \rightarrow 0$.  Some care needs to be taken with the zero-frequency terms, however.

The potential of a finite lattice may often be approximated by that due to an infinite lattice of charge.  This approximation will be reasonable provided the observation point is near enough to the lattice to feel negligible edge effects.  We may approximate the distance, $D$, where edge effects start to become significant by setting the magnitude of the dominant oscillatory term of the infinite lattice's potential equal to the one over the distance to the edge of the distribution.  If the characteristic spacing between charges is taken to be one and the projected distance to the edge is $L$, we find for one and two-dimensional Coulomb systems,
\begin{eqnarray}
D \sim \log L.
\end{eqnarray}
Numerically, it has been observed that as the observation point moves away from the charge distribution, there is often a relatively sharp transition region of distances over which the edge effects start to become significant.  A plot demonstrating this transition for a linear array of Coulomb charges is shown in Fig.(\ref{fig:one}).  A fit to the linear region of this data shows that the slope is approximately equal to $-6.281$.  This is quite close to the value of $-2\pi$ predicted by Eq.(\ref{oned}) for an infinite line of point charges.  
For Coulomb systems, the transition distance $D$ appears to be well described by the logarithmic fit
\begin{eqnarray}
D \approx 0.89 +0.27 \log L,
\end{eqnarray}
for an equally spaced, one-dimensional array, and by
\begin{eqnarray}
D \approx 0.34 +0.22 \log L,
\end{eqnarray}
for a square lattice, two-dimensional array.

The potentials of lattices slightly perturbed away from their equilibrium positions may also often be approximated by the potentials of perfect, infinite lattices.  Such a perturbed lattice may be considered to be the sum of a perfect lattice plus a series of physical dipoles.  The exponentially decaying terms due to the perfect lattice will dominate those terms due to the dipoles over some range which depends upon the size of the displacements.  Outside of this range the dipoles will dominate and the error will go as a power of one over the distance to the distribution.  We conclude that for observation points near large, clean physical systems, the results obtained here for infinite, perfect lattices may often apply.

\begin{figurehere}
\begin{center}\scalebox{.9}{\includegraphics{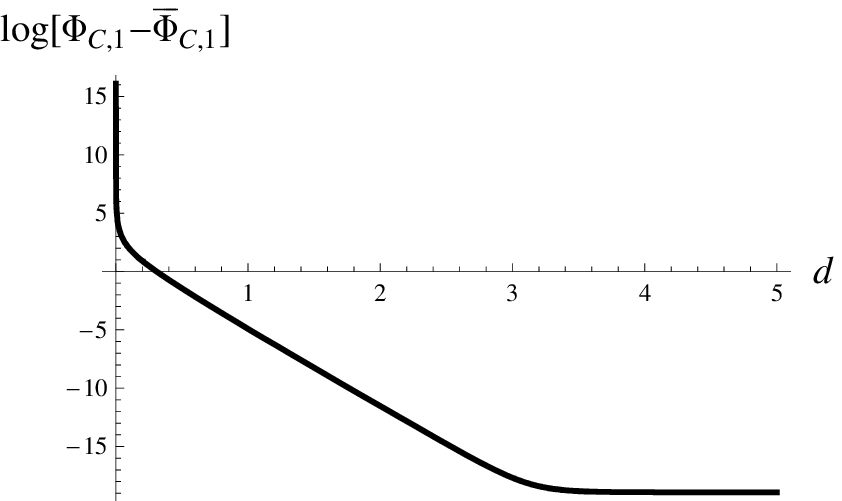}}
\caption{\label{fig:one} A plot of the logarithm of the difference between $\Phi_{C,1}(d,0,0)$ and $\bar{\Phi}_{C,1}(d,0,0)$ versus $d$ for a $2 \times 6400$ charge linear array placed on the $y$-axis and centered at the origin.}
\end{center}
\end{figurehere}

\appendix
\section{Two-dimensional expansion}
From Eq.(\ref{Poisson}), we have
\begin{eqnarray} \nonumber
\Phi_{C,2} &=& \sigma \sum_{\textbf{G}} \int \frac{\exp[i\textbf{G}\cdot \textbf{R}]}{|\textbf{r}-\textbf{R}|} R d\theta d R \\ \nonumber
&=&\bar{\Phi}_{C,2} +\sigma \sum_{\textbf{G}\not = \textbf{0}} \exp[-i \textbf{G}\cdot \textbf{r}] \int \frac{\exp[i\textbf{G}\cdot \textbf{R}]}{\sqrt{d^2 + R^2}} R d\theta d R  \\ \nonumber
&=&\bar{\Phi}_{C,2}+  2 \pi \sigma \sum_{\textbf{G}\not = \textbf{0}} \exp[-i \textbf{G}\cdot \textbf{r}] \int_0^{\infty} 
\frac{\mbox{J}_0(|\textbf{G}| R)}{\sqrt{d^2 + R^2}} R d R \\
&=& \bar{\Phi}_{C,2} + 2 \pi \sigma \sum_{\textbf{G} \not = \textbf{0}} \frac{1}{|\textbf{G}|} \exp[i \textbf{G}\cdot \textbf{r}-|\textbf{G}|d].
\end{eqnarray}
In the above, $\mbox{J}_0$ is a Bessel function of the first kind.
\section*{Acknowledgements}
The author thanks Professor Joseph Rudnick for helpful discussions.

\bibliographystyle{h-physrev3.bst}
\bibliography{refs}
\bibliographystyle{plain}
\end{multicols}

\end{document}